# Classical Newtonian Monte-Carlo Calculation of Ionization of Atomic Hydrogen by Protons in Energy Range 0.05 to 1.0 MeV

Seifeldin Dabbour, Skyler McMullen and F. E. Cecil
Department of Physics, Colorado School of Mines
Golden, Colorado 80401

## Abstract

We have designed a Euler approximation Monte-Carlo code for the calculation of the ionization of atomic hydrogen by protons with energies between 0.05 and 1.0 MeV. This code is used to calculate the total ionization cross sections σ(E), the angular distribution double differential cross sections $d^2\sigma/(d\Omega - dE)$ for electrons with energies of 150 eV ionized by 300 keV protons and the energy distribution $d^2\sigma/(d\Omega - dE)$ for the ionized electrons emitted at $10^o$ by protons of energy 300 keV. Our calculations are compared to experimental values of these quantities as well as independent calculations. We find excellent agreement of our calculated cross section and reasonable agreement of our calculated double differential cross section with independent calculations and measurements.

## Introduction

The correspondence principal is one of the bed rocks of the early years of modern physics and postulates that the results of quantum physics go over to those of classical physics in some appropriate limit, e.g. high quantum numbers, or of large numbers of quanta, or as Planck's constant h goes to zero. Ionization of electrons from ground state atoms by energetic projectiles is a phenomenon to which the correspondence principle should be applicable as the final states of the ionized electrons are in the continuum and consequently at very large quantum numbers.

We have examined the ionization of atomic hydrogen by energetic protons (.05 - 1.0 MeV) assuming classical Newtonian mechanics. Our approach to this phenomenon is hardly new. J.J. Thompson (1912) [1] discussed the "ionization by moving electrified particles". His theory is based on the assumptions:

1. "Cathode or positive rays when they pass through an atom repel or attract the corpuscles in it and thereby give to them kinetic energy"
2. "When the energy imparted to a corpuscle is greater than a certain definite value-the value required to ionize the atom- a corpuscle escapes from the atom and a free corpuscle and a positively charged atom are produced."

Since this is a classical three body problem an exact analytical solution was not possible and required an approximate numerical approach. Our calculations were carried out by numerically integrating the equations of motion in the Euler Monte Carlo approximation. Our approach is similar to that of Abrines and Percival [2] who calculated the total ionization cross section for incident proton energies of 50, 200 and 500 keV. Other published reports of total ionization cross sections measurements of atomic hydrogen by protons are summarized in Table 1. As noted in this table, there are only two reported measurements in the energy range of the present report, to wit .05 to 1.0 MeV. Consequently our calculated ionization cross sections are only compared to those of Hooper et al. [4] and Shah et al. [6] as presented in Fig. 6 below. In addition to the total ionization cross section, we have calculated double differential cross sections $d^2\sigma/ (d\Omega - dE)$ for the angular and energy distributions of the ionization electrons. Our measurements of these two double differential cross sections are compared to the results of Rudd [5]

## Our calculation

The geometry of the calculation is shown in Fig 1. The components are the nucleus N of the target hydrogen atom, the electron e in the first circular Bohr orbit of radius $r0 = 5.29\ 10^{-11}$ m with a speed $2.186\ 10^6$ m/s corresponding to an electron kinetic energy of 13.6 eV and the incoming proton projectile P, initially at a distance zP0 = -20 r0 below the x-y plane, with x, y coordinates each randomly distributed between -10 r0 and +10 r0. The plane of the electron orbit is randomly distributed between 0 and $2\pi$. A crucial parameter in the Monte Carlo calculation is the time step size per iteration. This was assigned a value dt = 0.025 r0/v0 = 6.047 $10^{-19}$ s, compared to classical electron orbit time of $2\pi$ r0/v0. = 1.52 $10^{-16}$ s. The total electrostatic force on each particle by the other two is calculated from the positions. From this force the accelerations were calculated and used to iterate the velocities and hence to new positions. The iterations continued typically for a time 8000 dt at which time the incoming proton was at a distance $10^{-9}$ m, roughly 100 Bohr radii above the initial plane of the electron for a proton energy of 75 keV. We should emphasize that a given calculation was repeated for +/- 50% variations of the initial parameters, xP0, yP0, zP0 and dt. We found the ionization cross sections independent of these variations. A given calculation at proton energy $E_p$ consisted of launching repeatedly a large number, $N_{beam}$, of protons, typically ($10^5$ – $10^6$), whose (x, y) coordinates in Fig.1 were randomly distributed over a 20 r0 x 20 r0 frame of area A = 400 $r0^2$ $m^2$ and counting the number $N_{ion}$ of ionized atoms. The cross section is then:

$$[1]\ \sigma = (N_{ion}\, A)/N_{beam}\ m^2$$

with statistical uncertainty:

$$[2]\ d\sigma = ((N_{ion})^{1/2}\, A)/N_{beam}\ m^2$$

Details of a typical calculation in which the atom is not ionized are shown in Figs 2 and 3. In this calculation the incoming proton had an energy of 75 keV and initial coordinates (xP0, yP0, zP0) = (3 r0, 0, -20 r0). A plot of the position of the electron, proton and nucleus are shown in Fig. 2. In Fig 3, the kinetic, potential and total energies are plotted as a function of time. The electron orbit is perturbed between a time of about $3\times10^{-16}$ s as the proton approached the plane of the electron and 5 $10^{-16}$ s when the proton leaves the vicinity of the atom. The intense periodic nature of the kinetic and potential energies for times greater the $5\times10^{-16}$ s reflects the fact the electron is in an elliptical bound orbit. The fact that the total energy remains negative demonstrates that the atom is not ionized

Details of a typical calculation in which the atom is ionized are shown in Figs 4 and 5. In this calculation the incoming proton had an energy of 75 keV and initial coordinates (xP0, yP0, zP0) = (1.1 r0, 0, -20 r0). A plot of the position of the electron, proton and nucleus are shown in Fig. 4. In Fig 3, the kinetic, potential and total energies are plotted as a function of time. The electron orbit is perturbed for time greater than about 3 $10^{-16}$ s as the proton approaches the plane of the electron. The fact that the total energy of the electron remains positive after the proton passes the atom demonstrates that the atom is ionized

## Our results

The results of our calculations are given Figures 6, 7 and 8. Figure 6 compares our calculated cross section to the measurements of Hooper [4] Shah[6]. Hooper et al. present the results of their measurements not as cross sections at individual energies but as a parameterization (with uncertainties) of the cross section between energies of 0.15 – 1.10 MeV. We evaluated their parameterization at energies between of our calculated cross sections. In addition, we calculated the angular distribution (double differential cross sections $d^2\sigma/ (d\Omega - dE)$) for electrons with energies of 150 eV ionized by 300 keV protons by recording the angle of emission of the electrons. Figure 7 compares our calculated angular distribution to the measurements of Rudd [5]. Finally, we calculated the energy distribution $d^2\sigma/ (d\Omega - dE)$ of electrons emitted at an angle of $10^0$ by recording the electron energies for ionization in which electron angle was $10^0$, again for a proton bombarding energy of 300 keV. Figure 8 compares our calculated ejected electron energy distribution to the measurements of Rudd [5]. As noted earlier, our method is comparable to the calculation of total cross section by Abrines and Percival [2]. Their results are compared to ours in Figure 9 where it is noted that our results exceed theirs by about 20%. The results of Hooper et al [4] and Rudd et al [5] assumed a target of diatomic hydrogen $H_2$ while our calculation assumed monatomic hydrogen. The results of Hooper et al [4] and Rudd et al. [5] were reduced by roughly a factor of two according to the procedure described by Hooper at al. [4].

# Our conclusions

The excellent agreement between our calculations of total ionization cross section, as presented in Figure 6, and the measurements of Hooper et al [4] and Shah[6] for energies between about 100 keV and 1.0 MeV support our original hypothesis that the correspondence principle is applicable to reaction cross sections. Likewise there is relatively good qualitative agreement between our calculation of the angular distribution of the ionized electrons and the measurements of Rudd [5] as indicated in Fig. 7 for angles between about 0.1 and 1.2 radians. Specifically, while the maxima in both distributions results are at about $45^o$, the magnitude of our calculated maximum is about 30% below the measured value of Rudd [5]. There is likewise very good agreement between our calculated electron energy distribution and that of Rudd [5] as seen in Fig. 8 for electron energies between 200 eV and 800 eV. A very appealing aspect of our approach is its simplicity, particularly in light of the fact that the system is three body and hence analytically unsolvable. Moreover, the calculation is quite transparent; a copy of the complete Mathematica code for a proton bombarding of 10 keV is attached as an appendix to this article. In this calculation we run the code for 200 incident protons. We record 15 ionizations for an ionization cross section of (8.29 +/- 2.16) x $10^{-16}$ $cm^2$.

Given the encouraging results of the present investigation, we expect to apply a modified version of the present code to other phenomena. Specifically, the ionization of atomic hydrogen by alpha particles could be studied by simply replacing the incoming projectile protons by alpha particles. The calculated total cross sections and differential cross sections could then be compared to the measurements of Puckett et al [6]. In addition, we could calculate the charge exchange cross sections for protons or alpha particles by counting those ionization encounters in which the ionized electron leaves the target hydrogen atom and is captured by the projectiles as determined by the final separation between projectile and electron remains relatively small (< 10 r0?) as the distance between the target and projectile becomes very large (>$10^3$ r0 - $10^4$ r0). Again, these calculations can be compared to measured proton-Hydrogen charge exchange cross sections [7].

Table 1

Published low energy ionization cross section measurements of atomic hydrogen by protons

| **Reference** | **Energy range** |
|---|---|
| Fite, Wade L. et al., Phys. Rev. **119**, 663 (1960) | 400 to 40,000 eV. |
| J. W. Hooper et al., Phys. Rev. **121**, 1123 (1961) | .15 to 1.10 MeV |
| H.B. Gilbody et al. Proc. Royal Soc.A, **277**, 137 (1964) | 60 to 400 keV |
| Shah, M.B. et al., J. Phys. B **14** ,2361 (1981) | 38 to 1500 keV |
| Shah M.B. *et al.,* J. Phys. B **20**, 2481 (1987) | 9 to 75 keV |
| Sahoo S et al. ,Phys. Rev. A **59**, 275 (1999) | 5 to 300 keV |
| Nasser T.E.I., Results in Physics, **17,** 103040, (2020) | 75 keV |
| Rudd M.E. et al., Phys. Rev. A **3**, 1635 (1971). | 50 to 300 keV |
| Solov'ev E.S. et al: Soviet Phys. JETP 18, 342 (1964). | 10 to 180 keV |

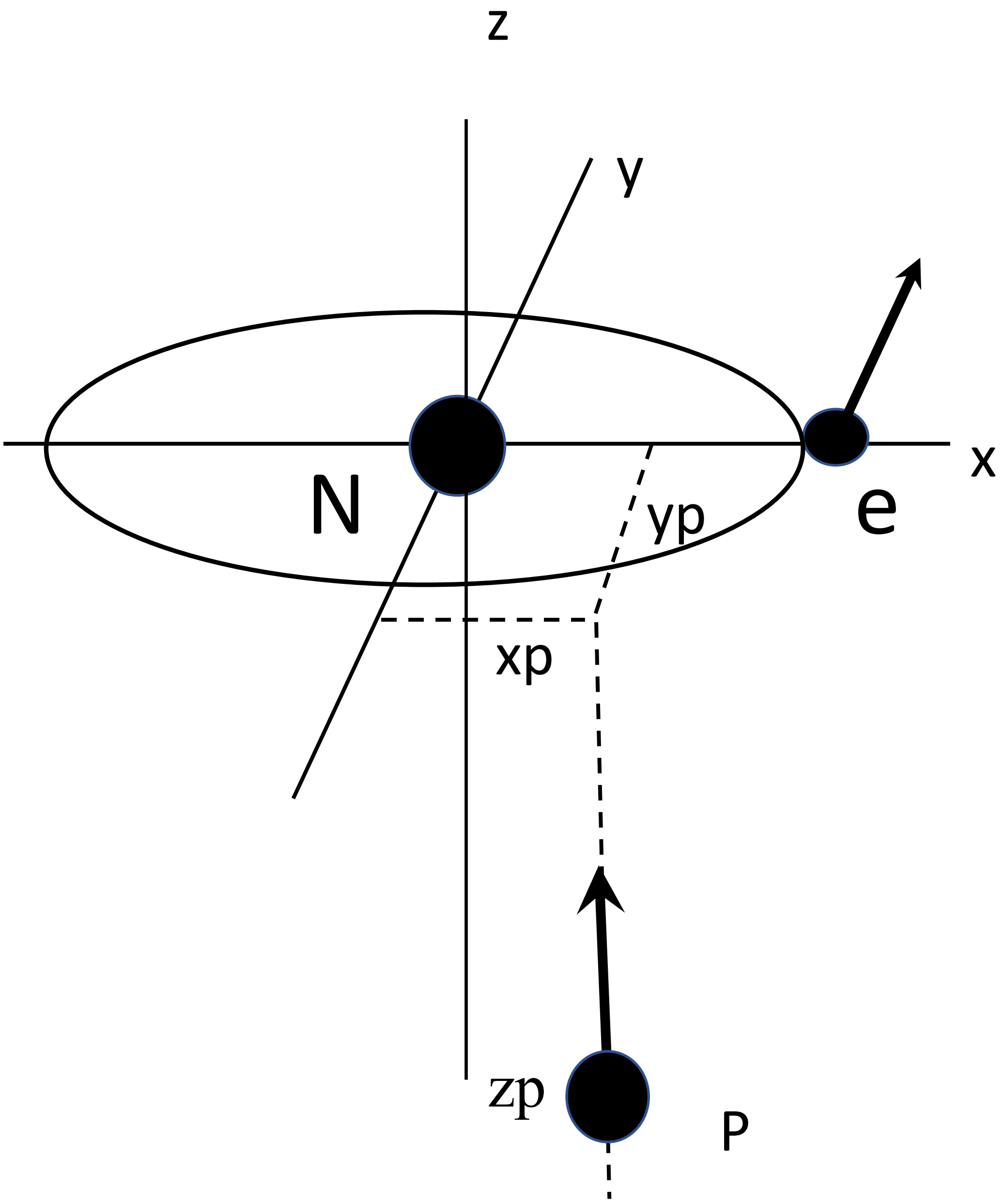

Figure 1. Geometry of the calculation showing the proton nucleus N of the target hydrogen atom, the atomic electron e in the first Bohr orbit of the atom with radius r0 = 5.29 $10^{-11}$ m and with a speed 2.186 $10^{6}$ m/s corresponding to an electron kinetic energy of 13.6 eV, and the incoming proton projectile P starting at a distance 20 r0 below the plane of the atomic electron with x,y coordinates randomly distributed over 10 r0 x 10 r0 frame.

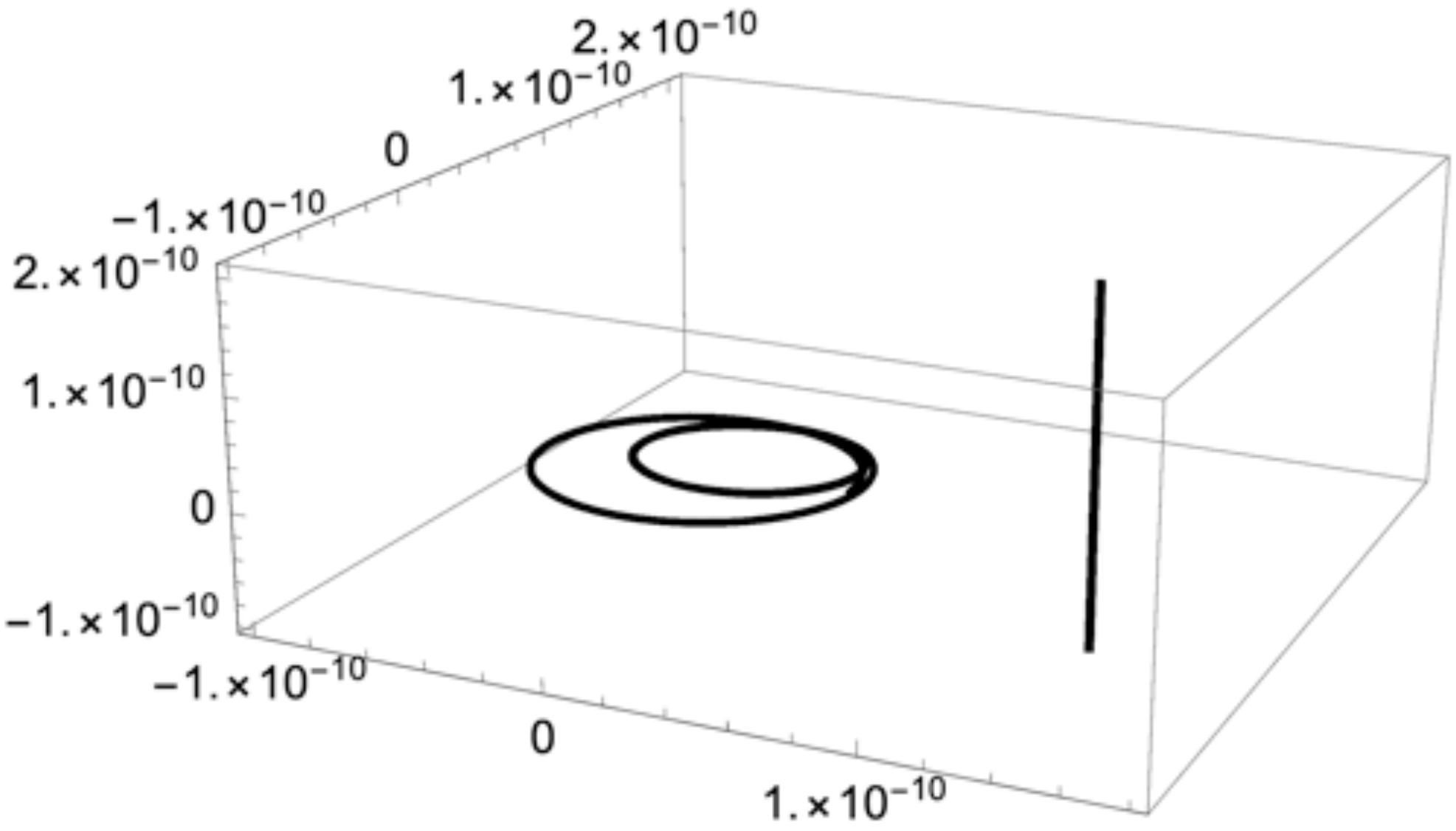

Figure 2. Trajectories of electron and proton for interaction where atom is not ionized. The initial position of the incoming proton is (3 r0, 0, -20 r0). The distances are in m.

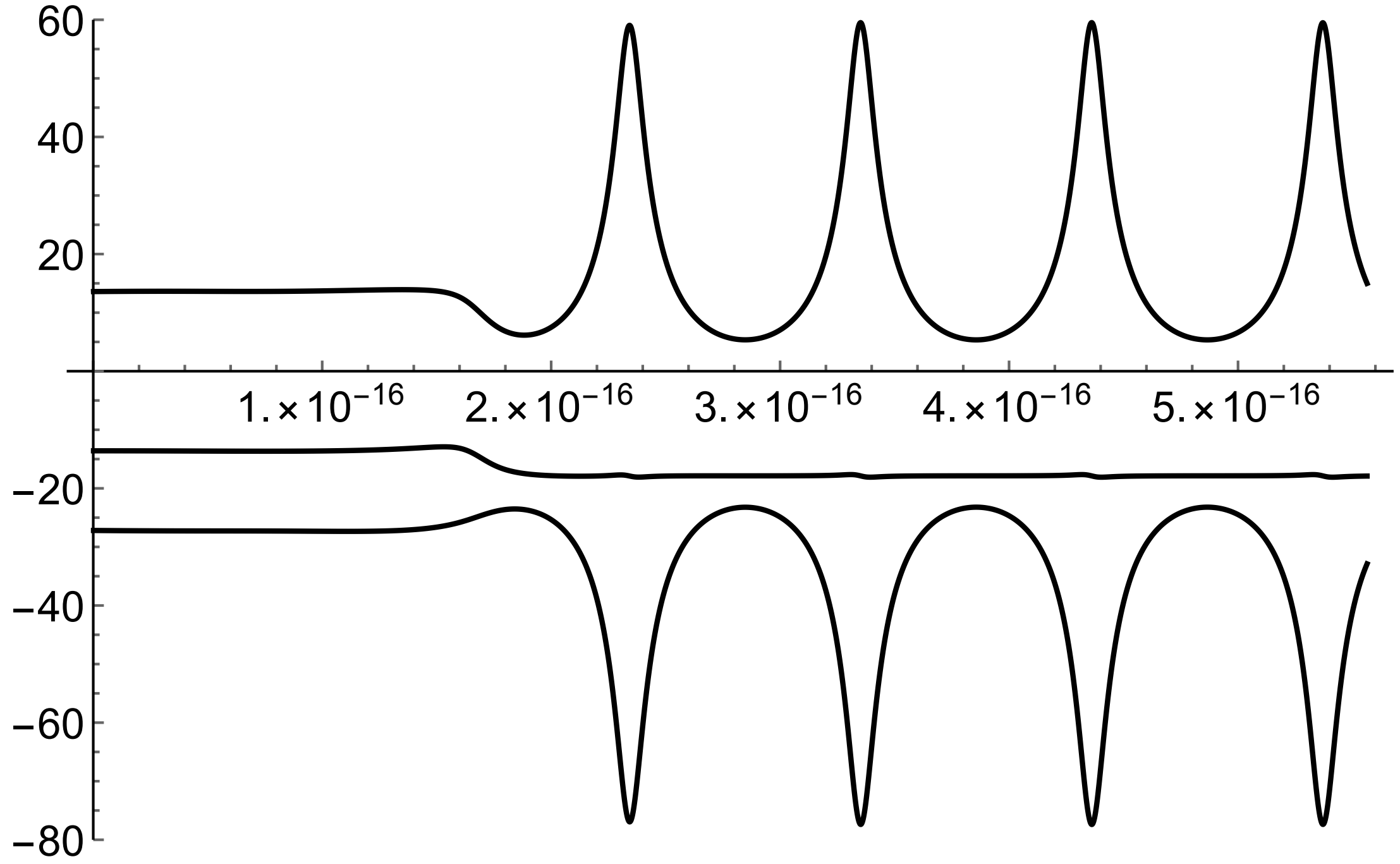

Figure 3. Energies of the electron (eV) as a function of time (s) after the proton is at a distance of 20 r0 below the plane of the electron. (Top=Kinetic, Middle=Total, Bottom=Potential) The fact that the total energy remains negative indicates the atom is not ionized. The striking time dependences of the kinetic and potential energies reflects the fact the incoming proton leaves the electron in a highly elliptical but un-ionized orbit.

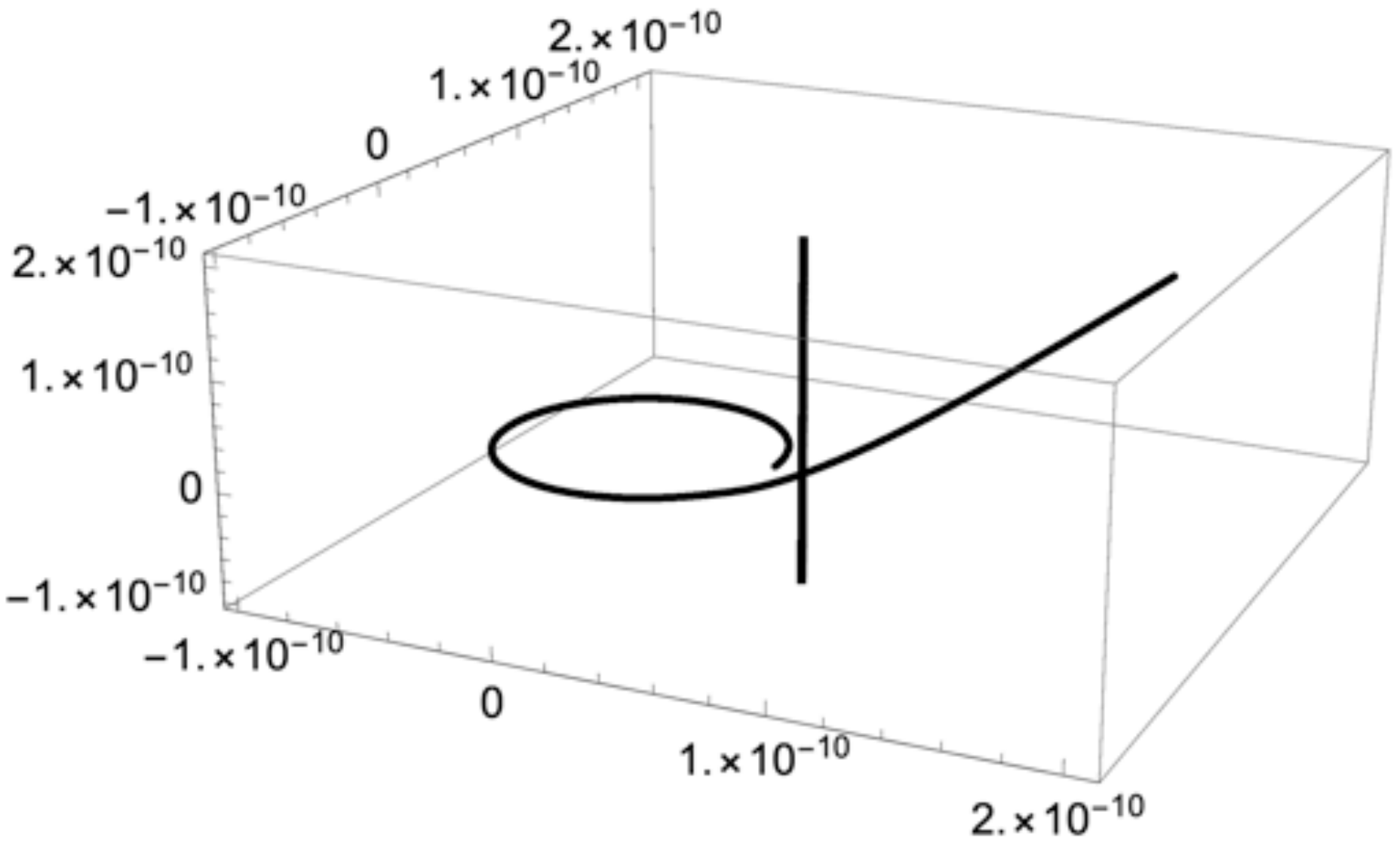

Figure 4. Trajectories of electron and proton for interaction where atom is ionized. The initial position of the incoming proton is (1.1 r0, 0, -20 r0). The distances are in m.

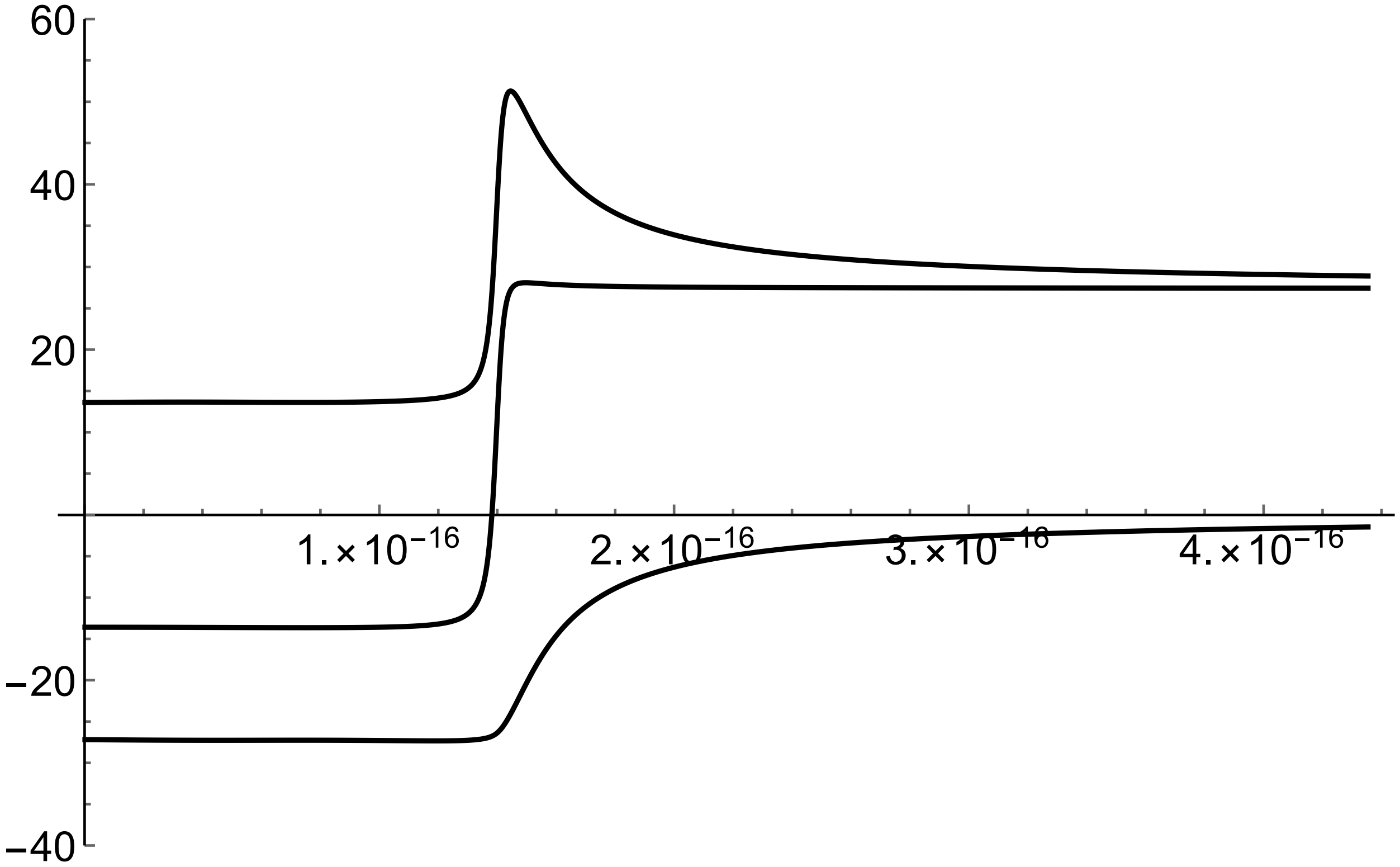

Figure 5. Energies of the electron (eV) as a function of time (s) after the proton is at a distance of 20 r0 below the plane of the electron. (Top=Kinetic, Middle=Total, Bottom=Potential) . The fact that the final total energy is positive indicates the atom is ionized.

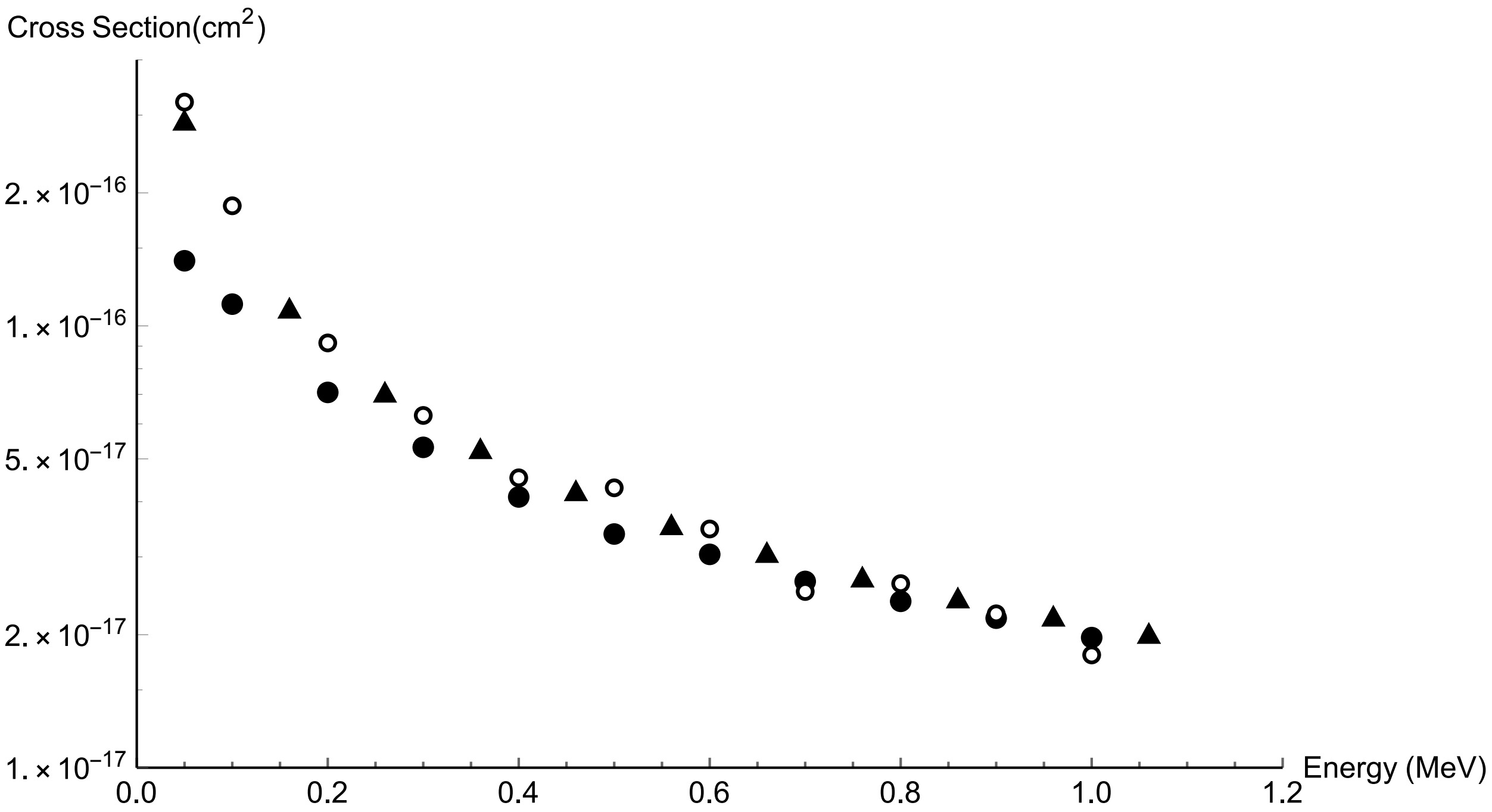

Figure 6. Comparison of measured and calculated values of total cross section for ionization of atomic hydrogen by protons . Filled Circles, Shah et al. [6], Triangles, Hooper et al.[4], Open Circles, present work. The error bars are the size of the plot symbols.

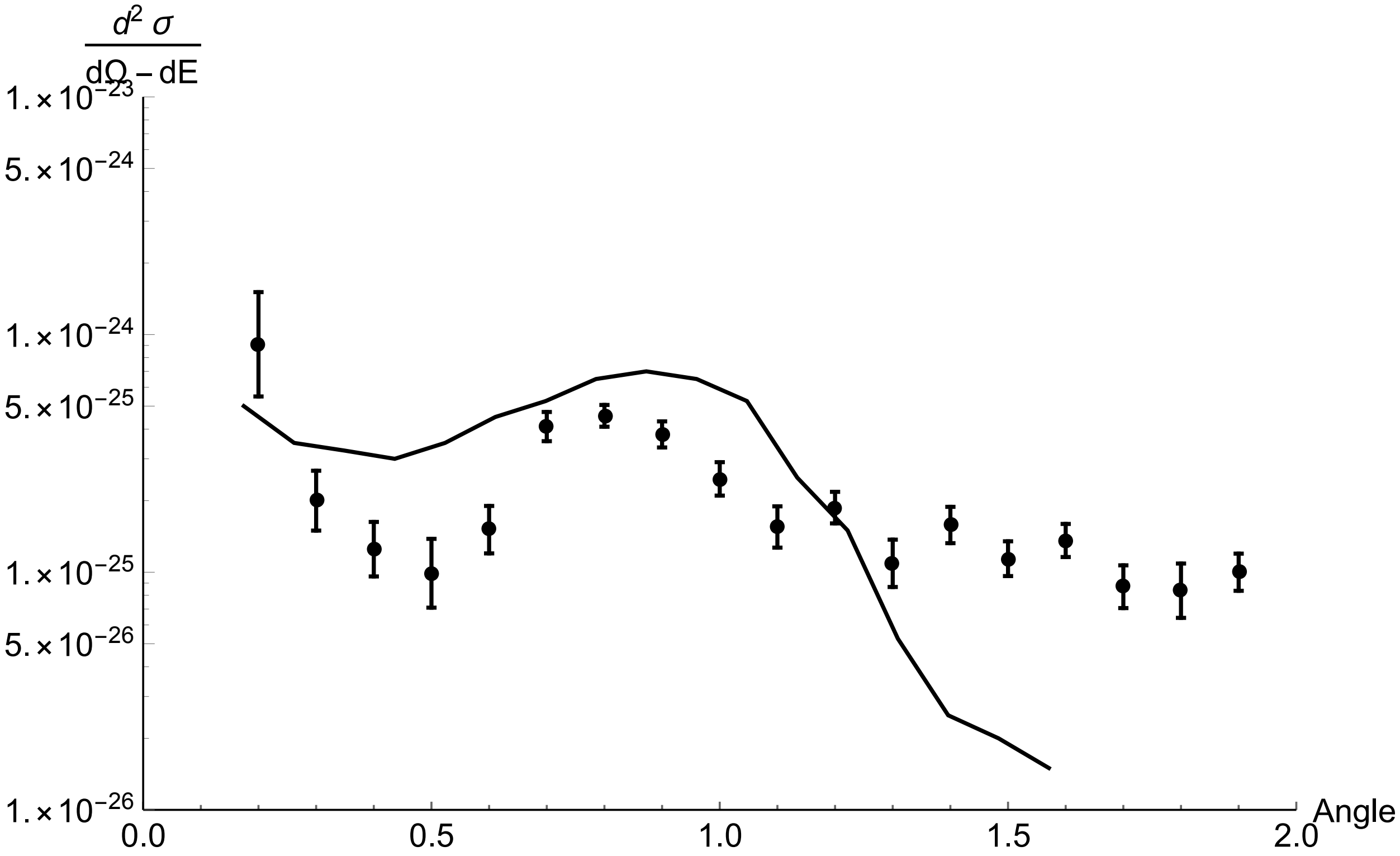

Figure 7, Angular distribution double differential cross section $d^2\sigma/(d\Omega - dE)$ $m^2/(eV- sr)$ for (150 +/- 25} eV electrons emitted during bombardment of monatomic hydrogen by 300 keV protons. Circles are present work; solid line Rudd et al. [5]. Angles are in radians.

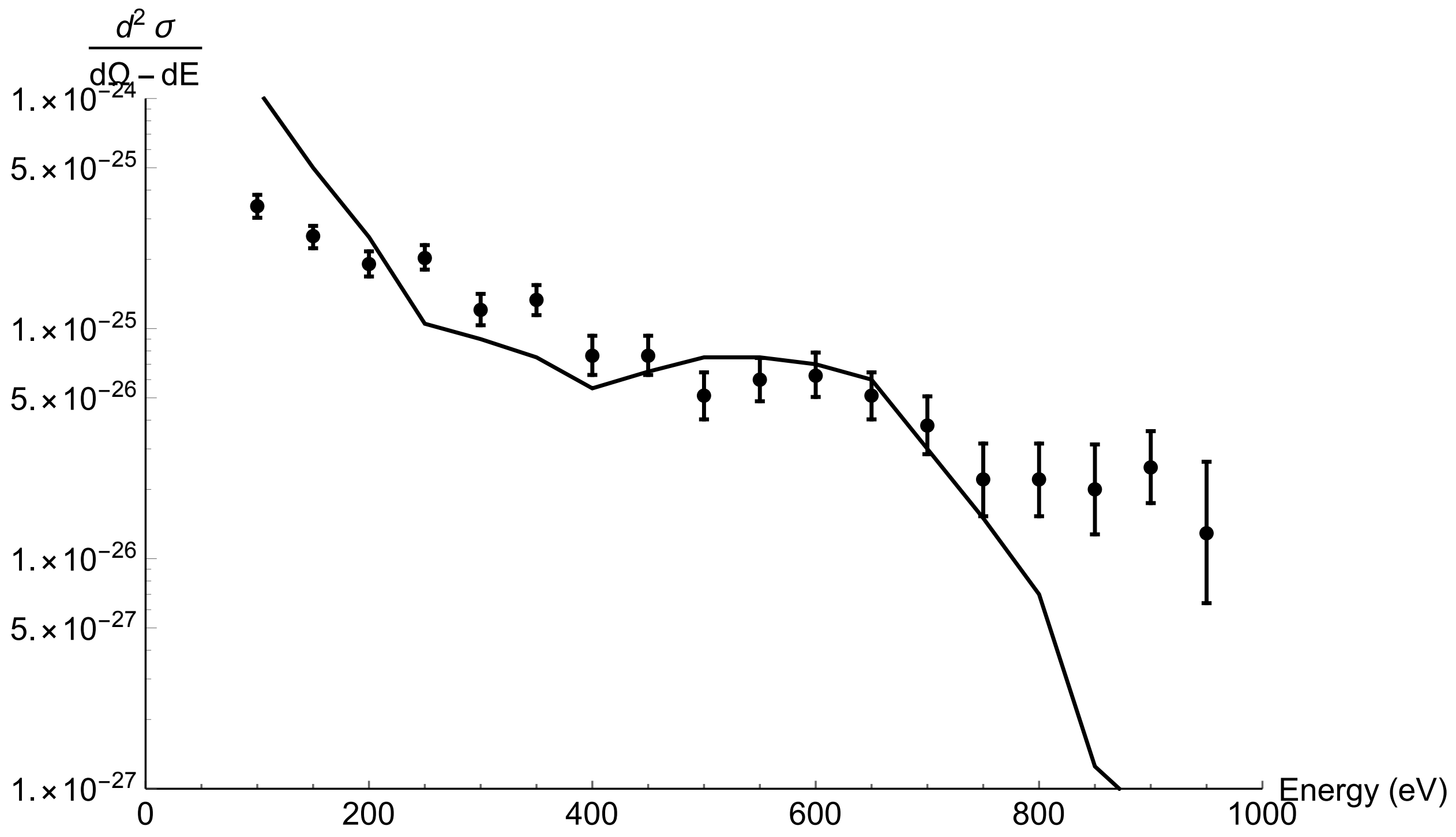

Figure 8. Energy distribution $d^2\sigma/ (d\Omega - dE)$ $m^2/(eV- sr)$ of electrons emitted at an angle of $(10 +/- 1)^0$ emitted during bombardment of monatomic hydrogen by 300 keV protons. Circles are present work; solid line Rudd et al. [5]

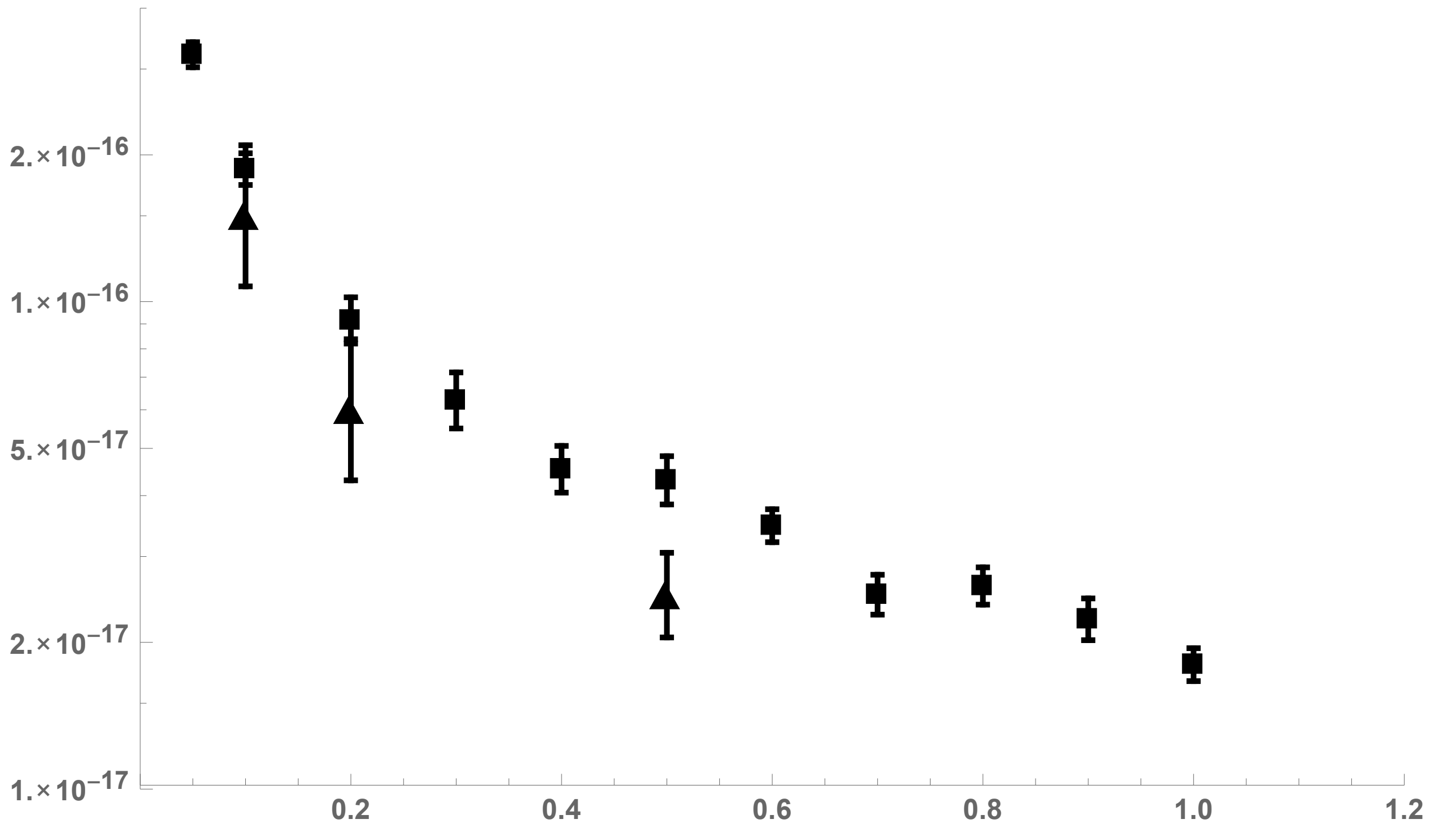

Figure 9. Comparison of our results (squares) to those of R. Abrines and I. C. Percival (triangles) [2].

**Appendix 1**

**Copy of MATHEMATICA code for 2000 incident protons at an energy of 10 keV**

```
eV = 1.6 10^-19; mE = 9.1 10^-31; mN = mP = 1.67 10^-27;
Ek = 13.6 eV; Ep = 10000 eV; r0 = 5.29 10^-11; t = 0;
v0 = Sqrt[(2 Ek)/mE]; vP = Sqrt[(2 Ep)/mP]; dt = .025 r0/v0;
k = 8.99 10^9; qE = -1.0 eV; qP = 1.0 eV; qN =1.0 eV;
 rP0 = r0; nbeam = 0; nhits = 0;
While[nbeam < 2000, {nbeam = nbeam + 1;
  xP = RandomReal[{-10 r0, 10 r0}]; yP = RandomReal[{-10 r0, 10 r0}];
  Θ= RandomReal[{0, 2 Pi}];
  zP = -20 r0; xN = yN = zN = 0; xE = r0; yE = 0; zE = 0;
  xE = r0 Cos[Θ]; yE = 0; zE = r0 Sin[Θ];
  vNx = vNy = vNz = 0; vPx = vPy = 0; vPz = vP; vEx = 0; vEy = v0;
  vEz = 0;
  rE = {xE, yE, zE}; rP = {xP, yP, zP}; rN = {xN, yN, zN};
  zP0 = zP/r0; xP0 = xP/r0; yP0 = yP/r0; t = 0;
  While[t < 2000 dt, {t = t + dt,
    rEN = Sqrt[(xE - xN)^2 + (yE - yN)^2 + (zE - zN)^2],
    rPN = Sqrt[(xP - xN)^2 + (yP - yN)^2 + (zP - zN)^2],
    rEP = Sqrt[(xE - xP)^2 + (yE - yP)^2 + (zE - zP)^2],
    FPN = (k qP qN)/rPN^2, FEN = (k qE qN)/rEN^2,
    FEP = (k qP qE)/rEP^2,
    FPNx = FPN ((xP - xN)/rPN), FPNy = FPN ((yP - yN)/rPN),
    FPNz = FPN ((zP - zN)/rPN), FNPx = -FPNx, FNPy = -FPNy,
    FNPz = -FPNz,
    FENx = FEN ((xE - xN)/rEN), FENy = FEN ((yE - yN)/rEN),
    FENz = FEN ((zE - zN)/rEN), FNEx = -FENx, FNEy = -FENy,
    FNEz = -FENz,
    FEPx = FEP ((xE - xP)/rEP), FEPy = FEP ((yE - yP)/rEP),
```

```
   FEPz = FEP ((zE - zP)/rEP), FPEx = -FEPx, FPEy = -FEPy,
   FPEz = -FEPz,
   aNx = (FNPx + FNEx)/mN, aNy = (FNPy + FNEy)/mN,
   aNz = (FNPz + FNEz)/mN,
   aPx = (FPNx + FPEx)/mP, aPy = (FPNy + FPEy)/mP,
   aPz = (FPNz + FPEz)/mN,
   aEx = (FEPx + FENx)/mE, aEy = (FEPy + FENy)/mE,
   aEz = (FEPz + FENz)/mE,
   vNx = vNx + aNx dt, vNy = vNy + aNy dt, vNz = vNz + aNz dt,
   xN = xN + vNx dt, yN = yN + vNy dt, zN = zN + vNz dt,
   vPx = vPx + aPx dt, vPy = vPy + aPy dt, vPz = vPz + aPz dt,
   xP = xP + vPx dt, yP = yP + vPy dt, zP = zP + vPz dt,
   vEx = vEx + aEx dt, vEy = vEy + aEy dt, vEz = vEz + aEz dt,
   xE = xE + vEx dt, yE = yE + vEy dt, zE = zE + vEz dt,
   kE = (mE (vEx^2 + vEy^2 + vEz^2))/(2 eV), uE = (k qE qP)/(eV rEN),
    TotE = kE + uE}],

  If[TotE > 0, nhits = nhits + 1, nhits = nhits],
  If[TotE > 0, Print[nbeam, " ", nhits]]}]
A = (20. r0 100)^2; Nbeam = 2000; Print[nhits, " ", nbeam, " ", \
zP/r0]
σ= nhits A/Nbeam
dσ = Sqrt[nhits] A/Nbeam
σ= 6.77215*10^-16
dσ= 6.1565*10^-17
```